\documentclass[twocolumn]{article}
\usepackage{epsf}
\usepackage{cdfnote}
\usepackage{physics_symbols}
\usepackage{dcolumn}
\input FEYNMAN

\newcolumntype{d}[1]{D{.}{.}{#1}}

\begin{document}
\runninghead{Present Searches for Higgs Signatures at the Tevatron}{Present 
Searches for Higgs Signatures at the Tevatron}
\normalsize\textlineskip
\thispagestyle{empty}
\setcounter{page}{1}

\twocolumn[
\vspace*{0.88truein}
\centerline{\bf PRESENT SEARCHES FOR HIGGS SIGNATURES AT THE
TEVATRON\footnotemark[1]}
\vspace*{0.37truein}
\centerline{L.~GROER}
\vspace*{0.015truein}
\centerline{\it Department of Physics and Astronomy}
\baselineskip=10pt
\centerline{\it Rutgers, The State University of New Jersey}
\baselineskip=10pt
\centerline{\it Piscataway, New Jersey 08855 USA}
\baselineskip=20pt
\centerline{\it for the}
\vspace*{0.015truein}
\centerline{CDF AND D\O\ COLLABORATIONS}
\vspace*{0.225truein}
\centerline{\footnotesize\today}
\vspace*{0.21truein} 

\abstracts{We present results for various searches for signatures of standard
and non-standard model Higgs boson decays conducted at the collider detectors
CDF and D\O\ using $\approx 100~\invpb$ of integrated luminosity each from the
Tevatron collider Run~1 (1992-96) at $\sqrt{s} = 1.8$ TeV\@.  No evidence for a
Higgs boson decay is found and various limits are set.}{}{}

\vspace*{12pt}
]

\footnotetext[1]{Submitted to Hadron Collider Physics XII, SUNY Stony Brook,
June 5-11, 1997.}       

\setcounter{footnote}{0}
\renewcommand{\thefootnote}{\alph{footnote}}

\section{Introduction} 

\noindent  Despite the enormous success of the standard model (SM) of particle
physics, the mechanism of electroweak symmetry breaking is still unknown.  The
most popular mechanism to induce this symmetry breaking is spontaneous symmetry
breaking of the SU(2) $\otimes$ U(1) electroweak gauge theory which results in
the gauge bosons and fermions acquiring mass via the Higgs
mechanism.\cite{gunion}
Another possible mechanism introduces dynamical symmetry breaking via 
technicolor interactions.\cite{eichten:1996}  Both of these mechanisms predict
the existence of a neutral heavy scalar ($X^0$) of unknown mass, which could be
produced at the Tevatron through $p\pbar \to W^\pm/Z^0 + X^0$ with a production
cross-section of the order of 0.1 pb to 10 pb.  Various searches have been
conducted at both CDF and D\O\ for this neutral scalar produced in association
with a heavy gauge vector boson.

Extensions to the SM also predict a charged Higgs boson ($H^\pm$) which, if
lighter than the top mass, could allow large ${\cal B}(t \to H^+b$) which would
compete with SM predictions of ${\cal B}(t \to Wb$) $\approx 1$. CDF has
conducted searches for these decays.                                           

The results reported here come from both collider detectors at the Tevatron,
CDF and D\O, where each collected about 100 \invpb\ of integrated luminosity
during Run~1 (1992--96) from $p\pbar$ collisions at $\sqrt{s} = 1.8$ TeV.

\section{Neutral Heavy Scalars ($X^0$)}

\begin{figure}
\begin{picture}(8000,15000)
\THICKLINES
\drawline\fermion[\NE\REG](0,2000)[5000]
\drawarrow[\NE\ATBASE](\pmidx,\pmidy)
\global\advance\pfrontx by -1000
\put(\pfrontx,\pfronty){$q$}
\drawline\fermion[\NW\REG](\pbackx,\pbacky)[5000]
\drawarrow[\NW\ATBASE](\pmidx,\pmidy)
\global\advance\pbackx by -1000
\put(\pbackx,\pbacky){$\bar{q}$}
\drawline\photon[\E\REG](\pfrontx,\pfronty)[6]
\global\advance\pmidy by 1000
\global\advance\pmidx by -1000
\put(\pmidx,\pmidy){$W^*/Z^*$}
\drawline\photon[\NE\REG](\pbackx,\pbacky)[6]
\global\advance\pmidx by -2000
\global\advance\pmidy by 1000
\put(\pmidx,\pmidy){$W/Z$}
\drawline\fermion[\NE\REG](\pbackx,\pbacky)[2500]
\drawarrow[\NE\ATBASE](\pmidx,\pmidy)
\global\advance\pbackx by 1000
\put(\pbackx,\pbacky){$q,\ell/q,\ell^-$}
\drawline\fermion[\E\REG](\pfrontx,\pfronty)[2500]
\drawarrow[\W\ATBASE](\pmidx,\pmidy)
\global\advance\pbackx by 500
\global\advance\pbacky by -1000
\put(\pbackx,\pbacky){$q',\nu/\bar{q},\ell^+$}
\global\gaplength=600
\drawline\scalar[\SE\REG](\photonfrontx,\photonfronty)[3]
\drawarrow[\SE\ATBASE](\pmidx,\pmidy)
\global\advance\pmidx by -2000
\global\advance\pmidy by -1000
\put(\pmidx,\pmidy){$H^0$}
\drawline\fermion[\SE\REG](\pbackx,\pbacky)[2500]
\drawarrow[\SE\ATBASE](\pmidx,\pmidy)
\global\advance\pbackx by 500
\put(\pbackx,\pbacky){$b$}
\drawline\fermion[\E\REG](\pfrontx,\pfronty)[2500]
\drawarrow[\W\ATBASE](\pmidx,\pmidy)
\global\advance\pmidx by 1000
\global\advance\pmidy by 500
\put(\pmidx,\pmidy){$\bar{b}$}
\end{picture}
\caption{Feynman diagram of $p\pbar \to H^0 + W^\pm/Z^0$ associated production
and subsequent decay.}
\label{higgs_feyn}
\end{figure}
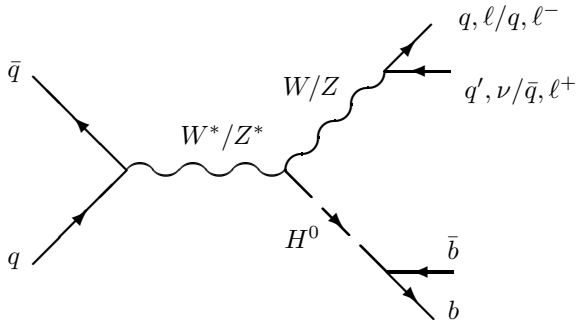

\noindent In the mass ranges that the Tevatron searches are sensitive to 
($M_{H^0} \approx 80$ -- $130~\gevcc$), the SM 
predicts\cite{gunion,stange:1994a,stange:1994c} that $H^0$ decays
preferentially to $b\bbar$.  The Feynman diagram for the associated production
processes, as well as the subsequent decay processes, searched for at the
Tevatron are shown in Figure~\ref{higgs_feyn}. Technicolor models
predict\cite{eichten:1996} that                                                 
$$\ppbar \to \rho_T^\pm \to W^\pm + \pi_T^0,\ \ \pi_T^0 \to  b\bbar$$
is also possible, where $\rho_T$ and $\pi_T^0$ are techni-mesons. 
The strategy therefore for the neutral heavy scalar searches is to look for
decay products of the associated vector boson and for $b$-jets from the scalar
boson decay. 

Both D\O\ and CDF search in the lepton + jets channels, looking for an isolated
lepton and missing \et\ (\met) to identify the $W^\pm \to \ell^\pm\nu$ decay,
and for jets, with $b$-tags, to identify the  $X^0 \to b\bbar$ decays.  CDF also
searches in the multijets channel, trying to identify $W/Z \to qq$ and $X^0
\to b\bbar$ decays.   D\O\ has a search looking for ``bosonic'' or
``fermiophobic'' Higgs ($H_F$) produced in association with a $W$ or $Z$
decaying to $qq$ and  $H_F$ decaying to $\gamma\gamma$ (discussed in section
3).

The SM expectations for the production cross-section $\sigma(\ppbar \to
W^\pm/Z^0 + H^0)$ at the Tevatron is less than 1 pb.\cite{stange:1994c}
As typical signal efficiencies in these searches are
less than 2\%, the current data set at the Tevatron is insufficient to set any
mass limits on the SM Higgs but new physics cross-sections of about a few
$\times$ 10 pb can be ruled out in the associated production channels.

\subsection{Search for $X^0$ in Lepton + Jets}

\noindent To identify the leptonic $W$ decay,  D\O\ looks for an isolated
electron (muon) with $\et \ge 25~(20)~\gev$ and $\met \ge 25~(20)~\gev$. To
identify the neutral scalar, two or more jets with $\pt > 15~\gevc$ are
required and one of the jets must be tagged with a muon of $\pt > 4~\gevc$
within a cone of $\Delta R< 0.5$. This helps tag the semi-leptonic decays of a
$b$-quark and cuts the QCD background.   Cleanup cuts are applied to remove $Z
\to \mu\mu$. Twenty-seven (12$e$, 15$\mu$) events remain from  100 \invpb\ of
Run~1 data.

Sources of background are $W$ with jets produced by gluon radiation ($13.9 \pm
2.1$), $t\tbar$ pair production ($7.2 \pm 2.5$), multijet events with false
lepton identification ($4.2 \pm 0.6$) and $Z \to \mu\mu$ + jets events ($0.2
\pm 0.1$).  Combined, these give $25.5 \pm 3.3$ background events expected,
which is consistent with the number observed.

The acceptance for identifying $WX^0$ events is calculated using Pythia
Monte-Carlo simulations of $p\pbar \to WH^0 \to \ell\nu + b\bbar$ under the
assumption of a neutral scalar with the  spin and decay properties of a SM
Higgs.  Acceptances for the combined lepton channels range from 0.52\% to
0.92\% for $M_{H^0}$ between 80 and 120 \gevcc\ with systematic uncertainties
of $\approx$ 8.3 (14.6)\% for electrons (muons). The acceptance includes ${\cal
B}(W \to e(\mu)\nu$).

\begin{figure}
{\epsfxsize=\hsize\epsfbox[125 465 270 610]{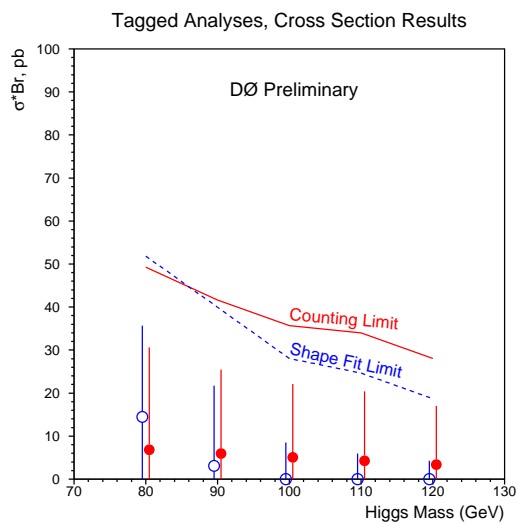}}
\caption{Fitted central values (filled dots - counting method, open dots -
shape fit) and 95\% C.L. upper limits on  $\sigma\cdot{\cal B}$ for SM $H^0$
associated production search at D\O.}
\label{d0_sigma_limits}
\end{figure}

A fit to the simple counting experiment, under the assumption of the SM $H^0$
decay, gives the $\sigma\cdot{\cal B}$ 95\% C.L. upper limit
shown in Figure~\ref{d0_sigma_limits}.

\begin{figure}
  {\epsfxsize=\hsize\epsfbox[125 465 270 610]{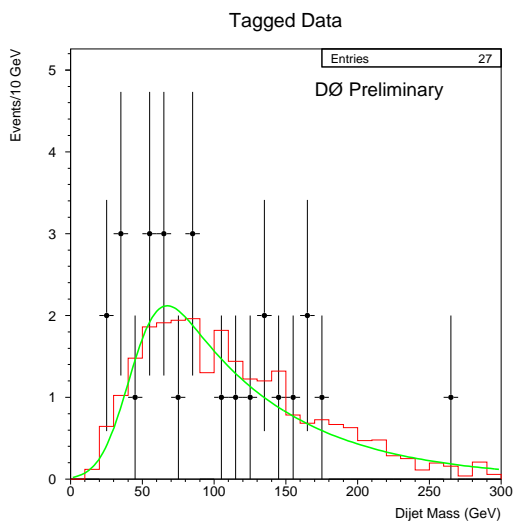}}
  \caption{D\O\ dijet invariant mass distribution for tagged data (points) and
           expected background (histogram).  The smooth curve represents the
           parameterized background.}
  \label{d0_dijet_mass}
\end{figure}

A slightly better limit can be set by using kinematic information in the data. 
Signal events should exhibit a peak in the dijet invariant mass spectrum.  No
mass peak above background is seen in Figure~\ref{d0_dijet_mass} which is a
plot comparing the 27 data points and the expected backgrounds.  A binned
maximum likelihood fit to the function \mbox{$\mu_i = N[f_i^B(1 - \alpha) +
f_i^S\alpha]$} is used to estimate the fraction, $\alpha$, of signal events
among the candidate 27 events.  The fraction of background (signal) in the 
$i^{\rm th}$ bin is represented by $f_i^B~(f_i^S)$.
This is done for a range of $M_{H^0}$ values
and the fit is consistent with zero in all cases.  Upper limits at 95\% C.L. 
are then set using the fitting technique.  The fitted and 95\% C.L. limits are
shown in Figure~\ref{d0_sigma_limits}.

CDF has performed a similar search looking for a central, isolated electron
(muon) with $\et~(\pt) > 20~\gev~(20~\gevc)$ and  $\met > 20~\gev$ to identify
the $W$ decay products.  Two jets only are required with $\et > 15~\gev$,
$|\eta| < 2.0$, to identify the scalar decay.  Either a single SVX $b$-tag (using
a secondary vertex found in the CDF Silicon Vertex detector) is required or a
double tag coming from two SVX tags or one SVX and one soft lepton tag (SLT)
which indicates a semileptonic $b$-decay.  A SLT is a track of $\pt > 2~\gevc$
associated with an electromagnetic calorimeter cluster or tracks in the muon
chambers (indicating an electron or muon) within $\Delta R < 0.4$ of a jet and
displaced from the primary vertex.  Various other cuts are applied to remove
$Z$'s, $\tau$'s and top quark candidates.

From 109 \invpb, 36 $(24e, 18\mu)$ single tag events and 6 $(2e,4\mu)$
double tag events remain.

Background events come predominantly from the direct production of $W$ bosons
in association with heavy quarks ($Wb\bbar$, $Wc\cbar$, $Wc$), mistags due to
track mismeasurements, and $t\tbar$ and single top production ($W^* \to tb$,
$gW \to tb$).\cite{singletop}  Other small backgrounds include $b\bbar$,
diboson ($WW$ or $WZ$) and Drell-Yan lepton pair production, and $Z \to
\tau\tau$ decays. $30 \pm 5$ single tag and $3.0 \pm 0.6$ double tag background
events are expected from all sources, which is consistent with the number
observed in the data.

Pythia 5.7 Monte-Carlo is again used to generate the  $p\pbar \to WH^0 \to
\ell\nu + b\bbar$ acceptance model for different $M_{H^0}$.  Acceptances range
from 0.53 to 1.1\% for single tags and from 0.17 to 0.42\% for double tags for
$M_{H^0}$ = 70 -- 120 \gevcc, including ${\cal B}(W \to e(\mu)\nu$)).
Systematic uncertainties on the acceptance are $ \approx 25\%$.

\begin{figure}
  {\epsfxsize=\hsize\epsfbox[55 145 540 620]{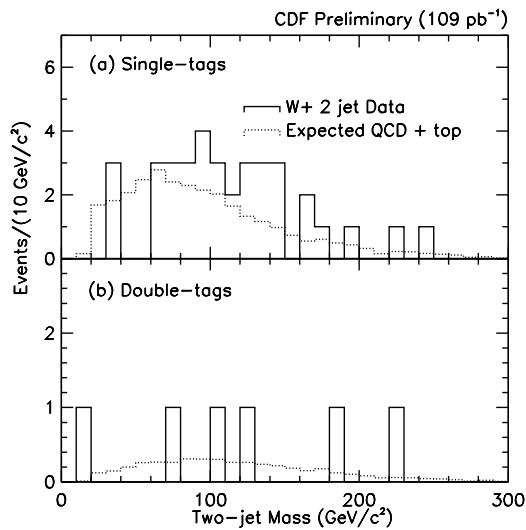}}
  \caption{CDF dijet invariant mass distribution for single and double tags.}
  \label{cdf_dijet_mass}
\end{figure}

Figure~\ref{cdf_dijet_mass} shows the dijet invariant mass distribution for the
single and double tagged events and the background expectations.  No mass peak
is seen and the slight excess is consistent with a background fluctuation at
the level of about one standard deviation.

\begin{figure}
  {\epsfxsize=\hsize\epsfbox[55 145 540 620]{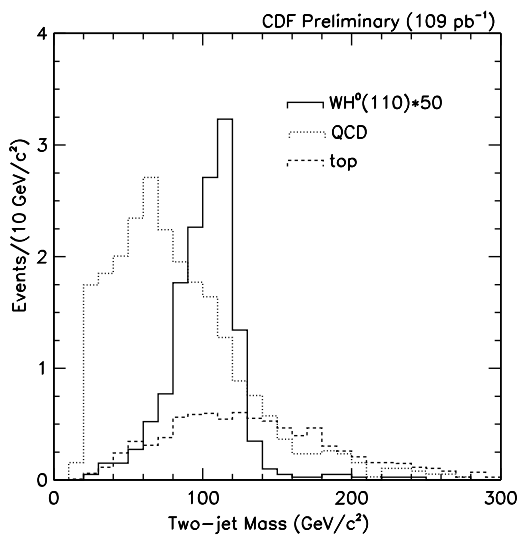}}
  \caption{Signal (for $M_{H^0}$ = 110 \gevcc) and background contributions to
 the fit to data for CDF search.  The $WH^0$ normalization scale has been blown
up by a factor of 50.}
  \label{cdf_fit_contributions}
\end{figure}

\begin{figure}
 {\epsfxsize=\hsize\epsfbox[0 0 455 495]{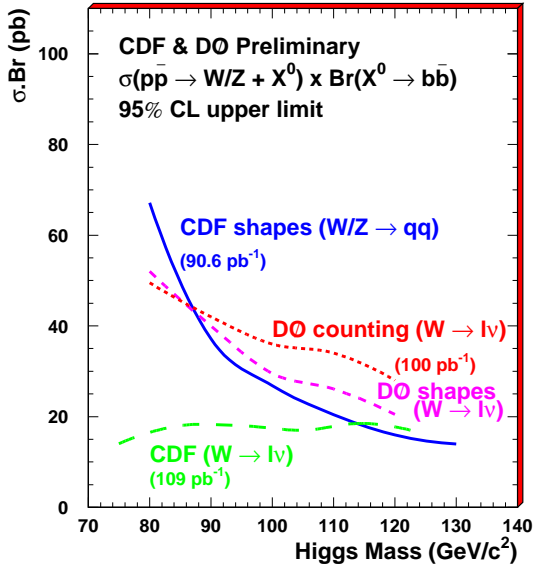}}
 \caption{Combined limits on new heavy neutral scalar production at the Tevatron
for Run~1.}
 \label{higgs0_limits}
\end{figure}

The number and shape of the dijet mass distribution are fit to a function 
$$\mu = f_{\scriptscriptstyle QCD}N_{\scriptscriptstyle QCD} +f_{top}N_{top} + 
f_{H^0}\cdot(\epsilon \cdot L \cdot \sigma \cdot {\cal B})$$
using a binned maximum likelihood method with the SM $H^0$ at various masses as
a model.   The contributions to the fit (shown in
Figure~\ref{cdf_fit_contributions} for a model with $M_{H^0}$ = 110 \gevcc) are
the expected backgrounds from QCD and top and the signal ($WH^0$). Fits for the
signal contribution are consistent with zero within about a standard deviation. 
Upper limits at 95\% C.L. can therfore be set for $\sigma\cdot{\cal B}$, shown
in the combined results plot, Figure~\ref{higgs0_limits}, as the CDF $(W \to
\ell\nu)$ line.  CDF excludes $\sigma\cdot{\cal B}$ $\gtsim 20$ pb, for 
$M_{H^0}$ = 75 -- 125 \gevcc.  This analysis has been submitted for
publication.\cite{abe:1997b}
                    
\subsection{Search for $X^0$ in Multijets}

\noindent CDF also conducts a search in the multijets channel for the
associated production where the vector bosons decay hadronically and the
neutral scalar decays to $b\bbar$.  The strategy here is to reconstruct the
$b\bbar$ invariant mass.  CDF selects events with four or more jets ($E_T \ge
15$ GeV, $|\eta| < 2.1$), two SVX $b$-tags and $p_T(b\bbar) \ge 50$ \gevc\
which cuts against QCD background but is still very efficient for the signal.
From 90.6 \invpb\ (the CDF Run~1B sample), 589 events remain.

Background events are calculated to be mainly QCD multijets with $b$-decays. 
From the fit described below, $453 \pm 25$ events are expected for $M_{H^0} =
100$ \gevcc.  The other main sources of background are events with fake
$b$-tags (estimated using tag rates from generic jet data to give $113 \pm 11$
events), $t\tbar$ production ($40.0\,_{ -\, 7.1}^{ + \, 8.1}$ events), $Z \to
b\bar{b}/c\bar{c}$ + jets ($12.5 \pm 5.1$ events) and $W/Z$ produced with heavy
flavor ($2.3 \pm 0.6$ events).

Signal acceptances are calculated from Pythia 5.6 Monte-Carlo assuming the
neutral scalar particle has the same spin and decay properties of the SM Higgs. 
Acceptances range between 0.6 and 1.7\% for $M_{H^0}$ in the range 70 to 140
\gevcc, including ${\cal B}(W/Z \to qq')$.  Systematic uncertainties are about
45\%.

A binned negative maximum likelihood fit is made to the shape of the observed
$b$-tagged dijet invariant mass spectrum, fitting a combination of signal, QCD
background, fake $b$-tags and top contributions with the function            
$$\mu_i\! =\! N^{data}[{\alpha N_i^{WH}} + 
{\beta N_i^{QCD}} + {\gamma N_i^{fakes}} +
{\omega N_i^{top}}].$$

\begin{figure}
{\epsfxsize=\hsize\epsfbox{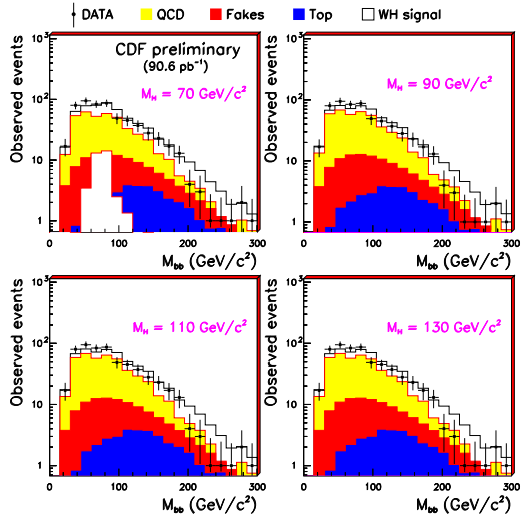}}
\caption{CDF $b$-tagged dijet invariant mass spectrum for multijets search. 
Fits for four different $M_{H^0}$ (70, 90, 110, 130~\gevcc) are shown.  
Histograms are not cumulative. The upper histogram shows the total for all 
contributions.}
\label{cdf_btag_dijet}
\end{figure}

Figure~\ref{cdf_btag_dijet} shows the fit contributions.  The QCD background
and $WH^0$ signal contributions are allowed to float in the fit.  The fitted
contribution for signal is consistent with zero for all $M_{H^0}$.  Upper
limits at 95\% C.L. can therefore be set for $\sigma\cdot{\cal B}$, shown in
the combined limit plot, Figure~\ref{higgs0_limits}, as CDF shapes ($W/Z \to
qq$).  As an example, this analysis excludes  $\sigma\cdot{\cal B} > 30~\pb$
for $M_{H^0} = 100~\gevcc$. 

\section{Search for ``Bosonic'' Higgs}\label{bosonic_sec}

\noindent ``Bosonic'' or ``fermiophobic'' Higgs ($H_F$) are predicted in some
non-minimal Higgs models.\cite{stange:1994c,diaz} Examples include two Higgs
doublet models (model I type) where only one of the doublets couples to
fermions and, more naturally, Higgs Triplet Models\cite{georgi} which
maintain $\rho = (M_W/M_Z cos\theta_W)^2 = 1$.  There are therefore no
tree-level couplings of $H_F$ to fermions and the $H_F$ decays preferentially
to the gauge vector bosons.  For $M_{H_F} \ltsim M_W$, $H_F$ decays
predominantly to two photons ($\gamma\gamma$) via vector boson loops. 
Technicolor predictions also include a $\pi^0_T \to  \gamma\gamma$
decay.\cite{eichten:1996} The
SM predicts that in this range ${\cal B}(H^0 \to \gamma\gamma) \simeq 0.001$
and $H_F$ does \emph{not} occur in the MSSM so this could be a very good
signature of new physics.  A bosonic Higgs with $M_{H_F} \ltsim 60$ \gevcc\ has
been ruled out by LEP~1.\cite{stange:1994c,acciarri:1996}

The production mechanism looked for at D\O\ is again associated production,
with $p\pbar \to W^\pm/Z^0 + H_F \to qq'/q\bar{q} + \gamma\gamma$, and an
attempt is made to reconstruct a mass bump of $M_{\gamma\gamma} > 60~\gevcc$
from the two photons.
             
D\O\ looks for two electromagnetic (EM) energy clusters with $\et$ of
$\gamma_1~(\gamma_2) > 20~(15)~ \gev$, and two jets with \et\ of $jet_1~(jet_2)
> 20~(15)~\gev$, with $40 < M_{jj} < 150$ \gevcc. Use of information in the
shower shapes of the EM energy clusters, the fraction of EM energy in the
calorimeters and requiring energy isolation and a clean ``tracking road'' where
there are no hits in the tracking chambers associated with the photon clusters,
all help to reduce the background and enhance the photon identification.
From 101 \invpb, 7 events remain, with none having $M_{\gamma\gamma} > 60$
\gevcc.  

\begin{figure}
  {\epsfxsize=\hsize\epsfbox[0 0 530 520]{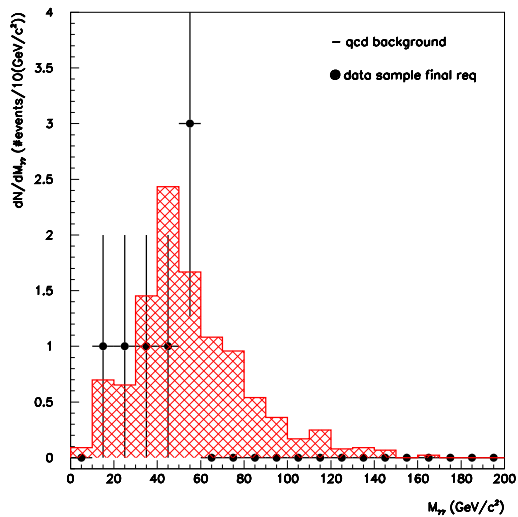}}
  \caption{D\O\ diphoton invariant mass spectrum.  No data events are in the 
signal region $M_{\gamma\gamma} > 60~\gevcc$.}
  \label{d0_bosonic_higgs_mgg}
\end{figure}

The largest background comes from QCD multijets where a jet fragments (with
probability $\sim 10^{-4}$) and
produces an isolated EM shower.  
The background is estimated from the data and normalized to 
the excluded $M_{\gamma\gamma} < 60$ \gevcc\ background region.
The expected background in the signal region $M_{\gamma\gamma} > 60$ \gevcc\ is
$3.5 \pm 1.3$ events.
Figure~\ref{d0_bosonic_higgs_mgg} shows the diphoton invariant
mass for the candidate events and the background.

Pythia is used to generate the acceptance model with  $p\pbar \to W/Z + H_F \to
qq'/q\bar{q} + \gamma\gamma$. Acceptances range from 5.5\% to 9\% for $M_{H_F}$
in the range 60 to 150 \gevcc.  Systematic and statistical uncertainties are
$\approx 10\%$.

\begin{figure}[t]
  {\epsfxsize=\hsize\epsfbox[0 0 520 520]{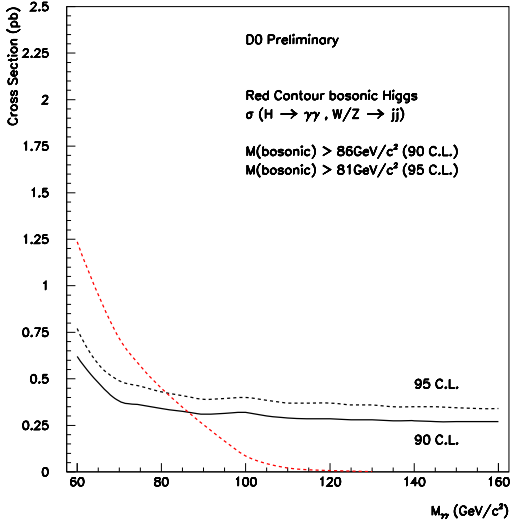}}
  \caption{D\O\ cross section limits for a bosonic Higgs boson decaying to
photon pairs (solid line at 90\% C.L. and dashed line at 95\% C.L.).
The dotted line is the theoretical calculation assuming SM
coupling strengths between $H_F$ and the weak gauge vector bosons.}
  \label{d0_bosonic_higgs_limit}
\end{figure}

There are no events in the signal region so 90\% and 95\% C.L. upper limits on 
$\sigma(p\pbar \to W/Z + X^0 \to jj + \gamma\gamma)$ are set, shown in
Figure~\ref{d0_bosonic_higgs_limit}.  New physics processes with $\sigma > 0.4$
pb are excluded at 95\% C.L..  Assuming SM coupling strengths between $H_F$ and
the weak gauge vector bosons,\cite{stange:1994c} $M_{H_F} > 81~(86)~\gevcc$ is
excluded at 95~(90)\% C.L..

\section{Charged Higgs ($H^\pm$)}

\begin{figure}
  {\epsfxsize=\hsize\epsfbox{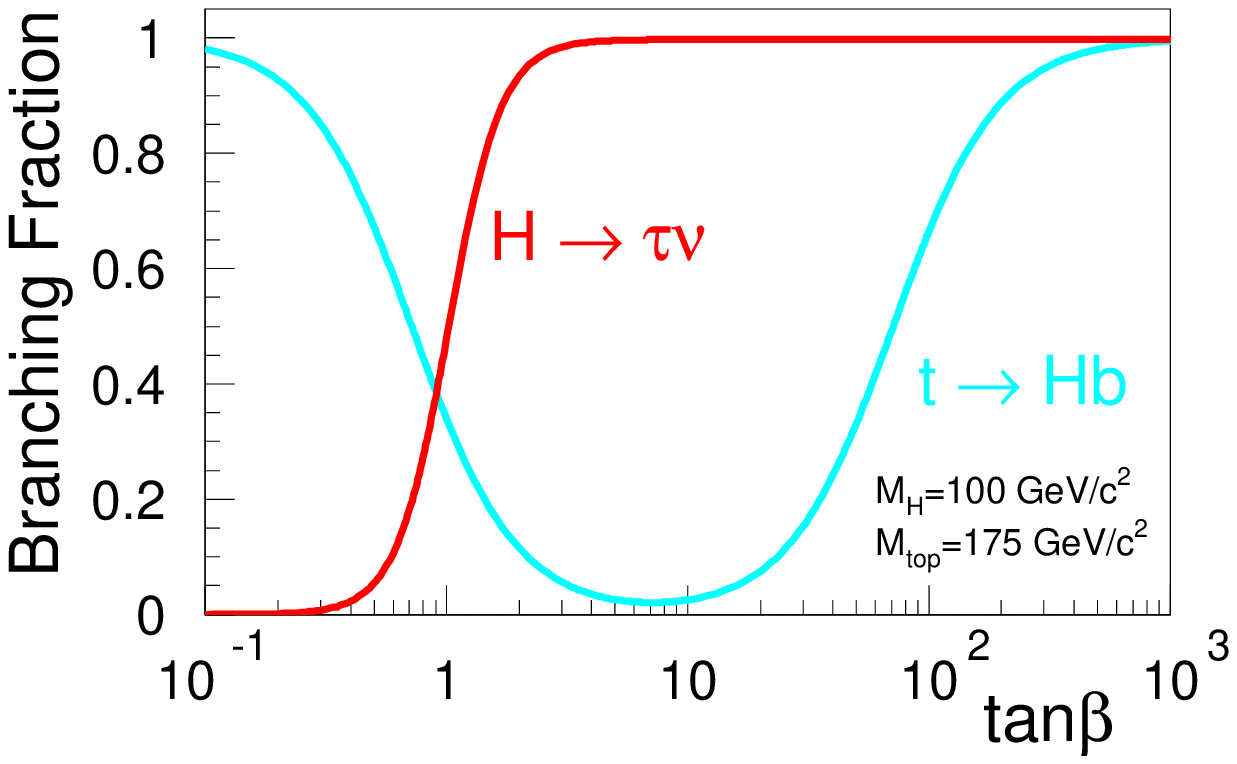}}
  \caption{Branching fractions for $t \to H^+b$ and $H^+ \to \tau^+\nu$ for
$M_{H^\pm} = 100~\gevcc$.}
  \label{ch_higgs_br}
\end{figure}

\noindent Many extensions to the SM, including supersymmetry and E(6) string
models, predict an extended Higgs sector with more then one physical Higgs
boson.\cite{gunion}  In particular, two SU(2) Higgs doublet models (model II
type) predict that one of the Higgs doublets couples preferentially to up-type
quarks and neutrinos and the other to down-type quarks and charged leptons. 
Five physical Higgs particles are predicted: two charged Higgs particles
($H^\pm$), two scalars  ($h^0$, $H^0$) and one pseudo-scalar ($A^0$).   The
predictions of these models depend on the parameters \tanb\ which is the ratio
of the vacuum expectation values of the two Higgs doublets, the mass of the
charged Higgs ($M_{H^\pm}$) and the top mass ($M_{top}$).  If  $M_{H^\pm} <
M_{top} - M_b$  then the top quark can decay to  $H^+b$ which competes with the
SM decay mode $W^+b$.  In these models, $H^+$ is predicted to decay
predominantly to either $\tau^+\nu$ or $c\sbar$. The decay branching ratios of
$t \to H^+b$ and $H^+ \to \tau^+\nu$ as a function of \tanb\ are shown in
Figure~\ref{ch_higgs_br}.

LEP 1~\cite{lep_results} has set model independent lower limits of $M_{H^\pm} >
44.1~\gevcc$ and CDF has published previous results\cite{abe:1996r} for large
\tanb\ values based on their Run~1A data sample of 19~\invpb.  LEP~2 is
expected\cite{duflot} to exclude charged Higgs bosons with $M_{H^\pm} <
52~\gevcc$.

CDF searches for $p\pbar \to t\tbar$ with at least one top decaying to $H^+b$,
directly for large values of \tanb\ ($\tanb > 10$) where $H^+ \to \tau^+\nu$
and indirectly for small values of \tanb\ ($\tanb \ltsim 1$) where $H^+ \to
c\sbar$.

\subsection{Direct Search for $H^\pm$ at High tan $\beta$}

\noindent CDF searches in events with large \met\ for hadronically decaying tau lepton candidates which are narrow
jets with one or three charged tracks.  A sliding \met\ isolation cut is used
to help identify the $H^\pm$ decay.  Two other jets and at
least one other lepton ($e$, $\mu$ or $\tau$) or jet is required.  All objects
are required to have $\et > 10~\gev$ with $\et > 20~\gev$ for the leading tau
candidate.  One or more SVX $b$-tags are required to identify the $b$-decay. To
gain acceptance for larger $M_{H^\pm}$ where the $b$-quark is produced with
less kinematic energy and falls below the \et\ cuts, events 
that have two hadronic taus ($\et > 30~\gev$) that are not back-to-back and 
that did not pass the previous selection are also selected. $Z \to
\ell\ell$ candidate events are removed.  From 100 pb$^{-1}$, 7 events remain (6
$\tau jjj$, 1 $\tau jje$ and zero from the ditau channel).

The largest source of backgrounds are fake taus, mostly coming from narrow QCD
jets with low track multiplicities.  Using fake rates measured from generic jet
data applied to the data sample, $5.4 \pm 1.5$ background events are expected
for both channels combined.  Electroweak processes such as $W/Z$ + jets and
diboson ($WW$, $WZ$, $ZZ$) production can be sources of real taus and
Monte-Carlo estimates give $2.0 \pm 1.6$ background events.  The total expected
backgrounds are $7.4 \pm 2.0$ events for both channels, which is consistent
with the number observed.

\begin{figure}[tp] 
  {\epsfxsize=\hsize\epsfbox{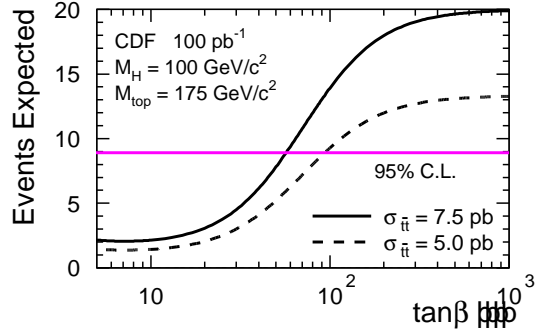}}
  \caption{Number of $H^\pm$ signal events expected as a function of $\tanb$
in CDF direct search for two different $t\tbar$ cross-sections 
(5.0 pb and 7.5 pb).  Models with $N_{H^\pm} > 8.9$ are excluded at 95\% C.L. }
  \label{ch_higgs_nexp} 
\end{figure}

Isajet 7.06 is used to generate $p\pbar \to t\tbar$ with the subsequent decay
of the top quarks to either $Hb$ or $Wb$ with $H \to \tau\nu$.
The acceptance, as a function of \tanb\ and $M_{H^\pm}$, is calculated by
convoluting the acceptance for the various decay modes with their expected
branching fractions.  Acceptances range between 3\% and 0.5\% for $M_{H^\pm}$ in
the range 100 to 165 \gevcc\ with a systematic uncertainty of 25\%. An example
of the number of expected signal events as a function of \tanb\ is shown in
Figure~\ref{ch_higgs_nexp} for $M_{H^\pm} = 100$ \gevcc\ and two different
values of the top production cross-section, $\sigma_{t\bar{t}} = 5.0~\pb$,
motivated by theory and $\sigma_{t\bar{t}} = 7.5~\pb$, chosen to be 50\% larger.
Using the acceptance model, the counting experiment can exclude at 95\% C.L.
all models that predict more than 8.9 signal events.

Figure~\ref{ch_higgs_final_limit} shows the $H^\pm$ mass limit as a function of
\tanb\ for $M_{top} = 175~\gevcc$ for the direct search and the indirect search
discussed in the following section.  At large values of \tanb, $M_{H^\pm} <
158~(147)~\gevcc$ is excluded for $\sigma_{t\bar{t}} = 7.5~(5.0)$ pb.

\begin{figure}[tp]
  {\epsfxsize=\hsize\epsfbox{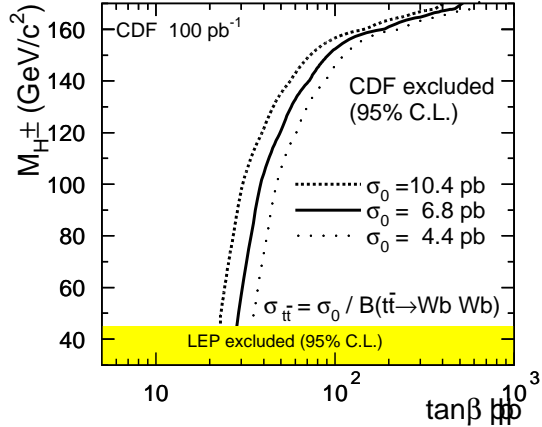}}
  \caption{CDF $H^\pm$ extended limit for large \tanb.}
  \label{ch_higgs_high_limit_with_top}
\end{figure}

This limit can be extended by including information from the top
discovery\cite{abe:1995l} by allowing $\sigma_{t\bar{t}}\cdot{\cal B}(t\bar{t}
\to W^+b\,W^-\bbar)$ to remain consistent with the CDF published result,
$\sigma_{obs}=6.8^{+3.6}_{-2.4}$~\pb.  As \tanb\ increases, $\sigma_{t\bar{t}}$
increases.  The resulting limits are shown in
Figure~\ref{ch_higgs_high_limit_with_top}.  This result has recently been
published.\cite{abe:1997a}

\vfill

\pagebreak

\subsection{Indirect Search for $H^\pm$ at \mbox{Low tan $\beta$}}

\noindent In regions of low \tanb, $H^+$ is predicted to decay predominantly to
$c\sbar$. A direct search in this channel is difficult due to the large QCD
backgrounds which obscure the $H^\pm$ decays so CDF rather considers the impact
a $H^\pm$ would have on the top dilepton and  lepton + jets search
channels.\cite{abe:1995l}

CDF searches for SM top decays ($t \to Wb$)  by looking for
the two leptonic decays of the $W$'s.  Events with two opposite-sign, high $p_T$
electrons or muons, $\met > 25~\gev$ and 2 jets ($\et > 10$ GeV, $|\eta| < 2.0$)
are selected. Nine events are observed with 2.2 $\pm$ 0.4 background events
expected in 109 \invpb\ from Run~1.

CDF also searches for one $W \to \ell\nu$ and one $W \to q\bar{q}^\prime$ by
selecting events with an isolated high-$p_T$  electron or muon, $\met > 20$ GeV
and three or more jets ($\et > 15$ GeV, $|\eta| < 2.0$).  At least one positive
SVX $b$-tag is also required to select against QCD backgrounds.  Thirty-four
events are observed with 10.4 $\pm$ 1.6 background events expected in 109
\invpb\ from Run~1.

Table~\ref{top_expect} show the number of events expected from the SM in each
channel for two different values of $\sigma_{t\bar{t}}$ (5.0 and 7.5 pb), after
background subtraction for non-$t\tbar$ processes.   The numbers observed are
consistent with SM expectations.  The backgrounds in each channel are slightly
higher than those in the standard top analyses, where the portion of the
background calculated from data is corrected for its top content under the
assumption that ${\cal B}(t \to Wb) = 1.0$.

\begin{table}
\begin{center}
\begin{tabular}{|c|c|c|}
\hline
                               &   {dilepton}      &  {lepton + jets} \\
\hline
 $\sigma_{t\bar{t}}$ = 5.0 pb  &  {4.1 $\pm$ 0.5}  &  {20.0 $\pm$ 3.0} \\
\hline
 $\sigma_{t\bar{t}}$ = 7.5 pb  &  {6.1 $\pm$ 0.7}  &  {30.0 $\pm$ 4.5} \\
\hline
 {\bf observed}                &  {\bf 6.8 $\pm$ 3.0} 
                                                   & {\bf 23.6 $\pm$ 6.0} \\
\hline
\end{tabular}
\end{center}
\caption{Expected number of signal SM $t\tbar$ events in 109 \invpb\ for CDF
dilepton and lepton + jets top search.}
\label{top_expect}
\end{table}

\begin{figure}
  {\epsfxsize=1.0\hsize\epsfbox[0 0 325 305]{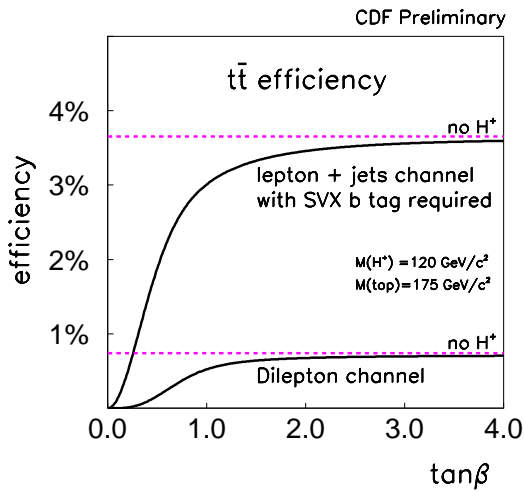}}
   \caption{Acceptance in the dilepton and lepton + jets channels at CDF for a 
charged Higgs model with $M_{H^\pm} = 120~\gevcc$.  The dotted lines represent
the acceptance for SM top decays in these channels.}
  \label{ch_higgs_eff}
\end{figure}

Pythia 5.7 is used to generate $t \to H^+b/W^+b \to c\sbar/\tau\nu$ for the
signal model.  Acceptance ranges from  0.01\% to 0.45\% for the dilepton channel
and 0.2\% to 2.4\% for the lepton + jets channel.  Systematic uncertainties are
about 12\% to 20\%.  Figure~\ref{ch_higgs_eff} shows the acceptance for both
channels as a function of \tanb\ for $M_{H^\pm} = 120~\gevcc$.  Note that the
acceptance drops very steeply towards low \tanb\ as here ${\cal B}(t\to H^+b$)
$\simeq$ 1 and $H^+ \to c\sbar$.  

If $H^\pm$ exists with low \tanb, then the production rates in the top dilepton
and lepton + jets channels would be much lower than the corresponding rates
predicted for SM top decays. Too many events have been observed in each of
these channels for ${\cal B}(t \to H^+b)$ to be very large.\footnote{For
example, only 1.4 dilepton and 12.3 leptons + jets events are expected for
$M_{H^\pm}$ = 120 \gevcc, \tanb\ = 0.6 and $\sigma_{t\bar{t}}$ = 5.0 pb.}
\ The channels are combined by simply summing the number of dilepton and lepton +
jets events. 

\begin{figure*}
  \hbox to \hsize{\centerline{%
  {\epsfxsize=0.5\hsize\epsfbox[0 0 325 305]{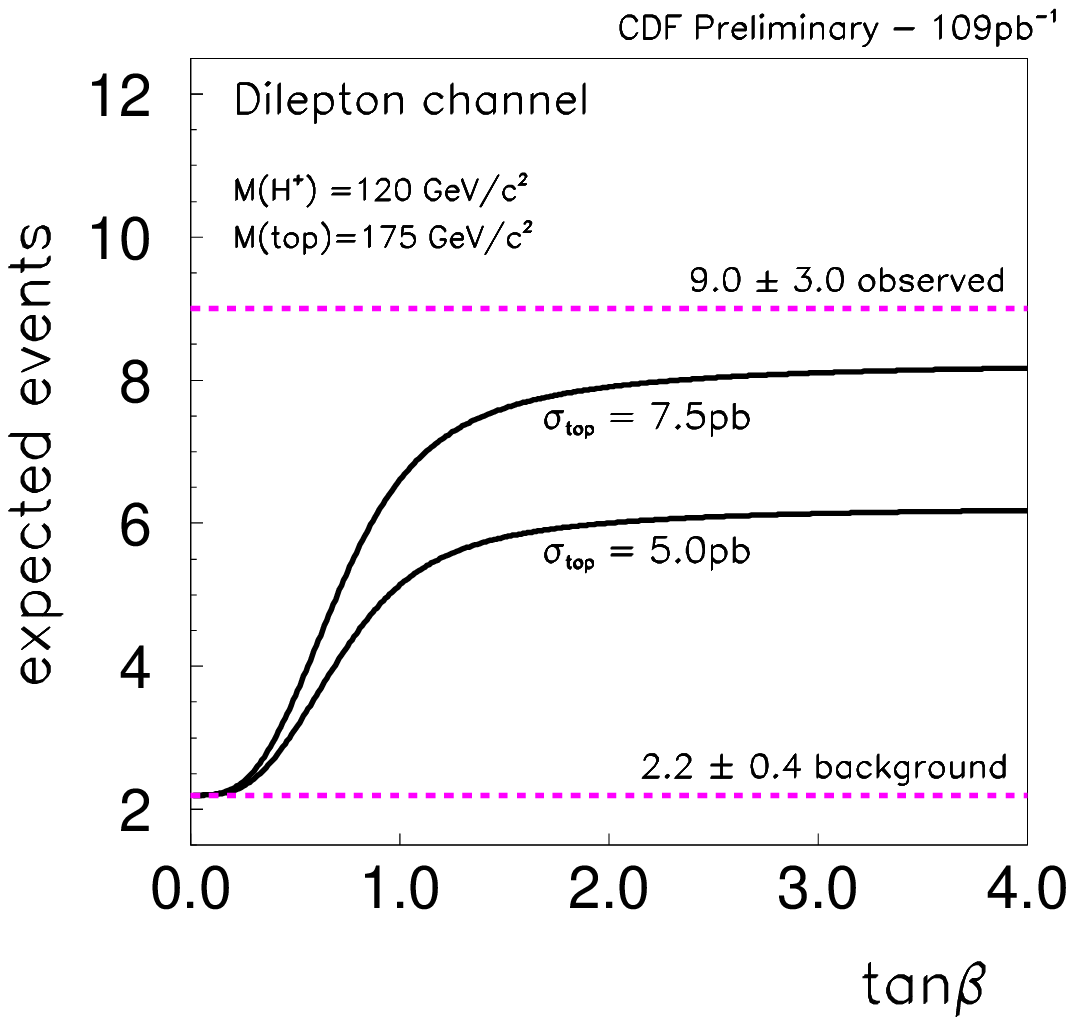}%
  \hfill%
   \epsfxsize=0.5\hsize\epsfbox[0 0 325 305]{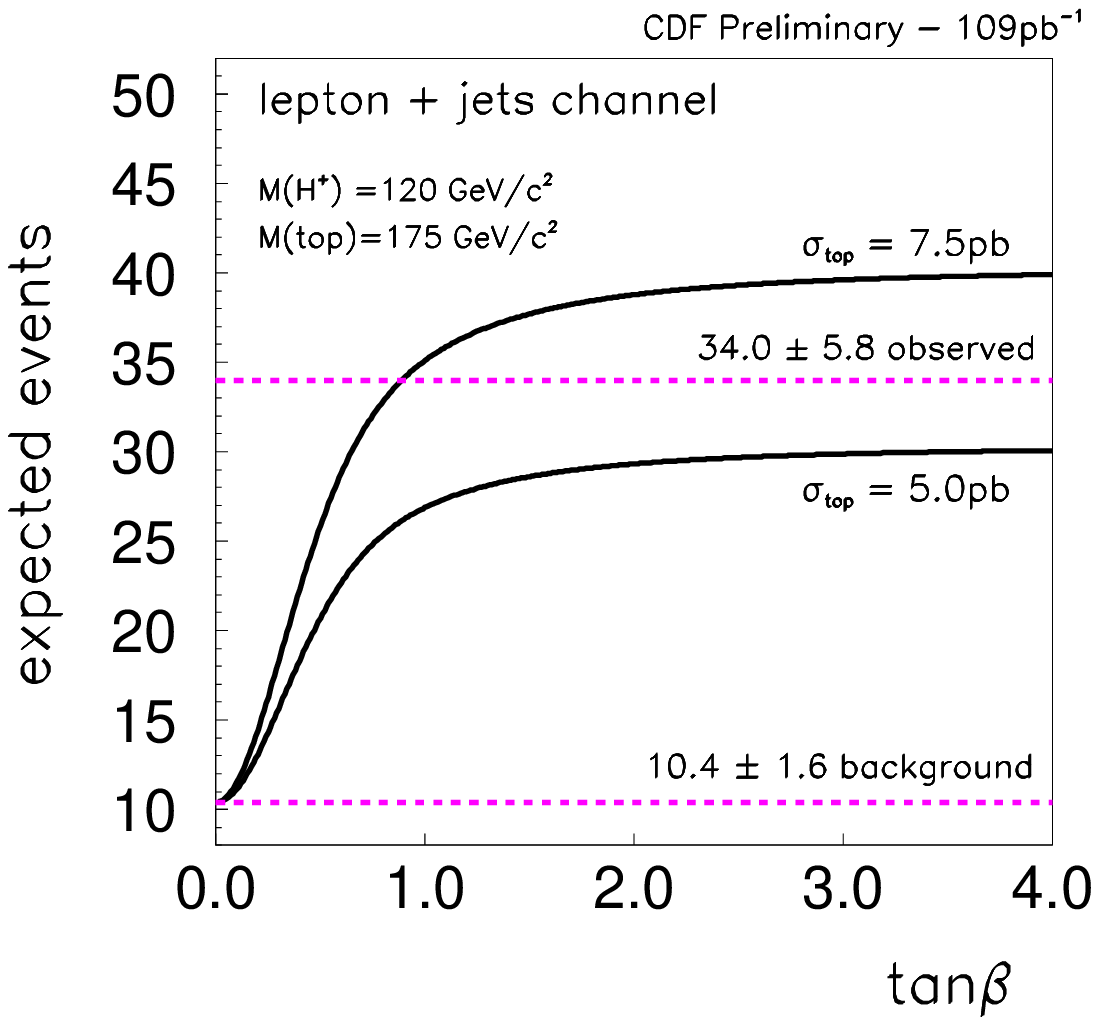}}}}
  \caption{Number of events expected at CDF in the top dilepton and leptons +
jets channels for a charged Higgs model with
$M_{H^\pm}$ = 120 \gevcc\ for $\sigma_{t\bar{t}}$ = 5.0 and 7.5 pb.
The number of events observed in the data and expected backgrounds are also
shown.}
  \label{ch_higgs_n_exp_low_tanb}
\end{figure*}

CDF calculates the number of expected events from the signal model
as a function of $M_{H^\pm}$ and \tanb\ and can exclude models which 
produce at least the number of observed events in less than 5\% of
pseudo-experiments.  The number of events expected for a model with $M_{H^\pm}$
= 120 \gevcc\  is plotted as a function of \tanb\ for both channels in
Figure~\ref{ch_higgs_n_exp_low_tanb}.

\begin{figure}
  {\epsfxsize=1.0\hsize\epsfbox{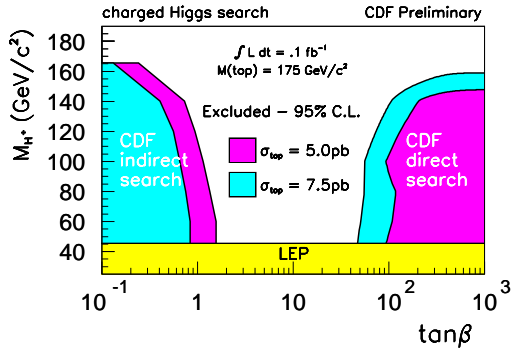}}
  \caption{CDF $H^\pm$ limits for direct and indirect search for two assumed
$\sigma_{t\bar{t}}$.} 
  \label{ch_higgs_final_limit}
\end{figure}

CDF excludes ${\cal B}(t \to H^+b) \gtsim 25~(50)\%$ for $\sigma_{t\bar{t}}$ = 
5.0~(7.5) \pb\ in the mass range $M_{H^\pm} = 60 - 165~\gevcc$ at 95\% C.L..
Mass limits for
charged Higgs at 95\% C.L. are shown in Figure~\ref{ch_higgs_final_limit} as a
function of \tanb\ for both CDF searches.

\section{Conclusions}

\noindent There is currently no evidence from the Tevatron collider Run~1 data
for any Higgs decays.  The Tevatron search limits are still about an order of
magnitude above the expected SM $H^0$ associated production cross-section in
$p\pbar$ collisions, therefore no mass  limits can be set.  However, limits on
any new physics processes in these channels have been set at about $\sigma >
20$ pb.  Limits on the associated production of bosonic higgs with $\sigma >
0.4$ pb and mass limits of $M_{H_F} > 81$ \gevcc\ at 95\% C.L. have also been
set.

Limits almost up to the top mass have been set on the mass of charged Higgs 
bosons for large and small values of \tanb, as well as branching fraction limits
for top quarks decaying via $H^\pm$.

The Tevatron collider Run 2 with center of mass energy $\sqrt{s} = 2.0$ TeV
and with instantaneous luminosities of  $\lum \sim 2 \times 10^{32}~{\rm
cm}^{-2}{\rm s}^{-1}$ is scheduled to begin in 1999.   Both detectors are
undergoing substantial upgrades.  D\O\ is currently installing a  silicon
vertex detector and solenoidal magnetic field and CDF is upgrading its  silicon
vertex system, which will enhance both their sensitivities to tagging $b$-quark
decays, crucial in searching for Higgs particles.\cite{amidei:1996} Both
detectors expect to take about 2 fb$^{-1}$ integrated luminosity over a couple
of years of running and sensitivities to $\sigma(p\pbar \to WH^0)$ of ${\cal
O}(1)$ pb are expected for intermediate mass Higgs of 80 -- 130 \gevcc.

Searches for non-minimal Higgs particles are also planned, looking for
couplings both to fermions and vector bosons pairs.  Studies indicate that we
might be sensitive to $H^+H^-$ pair production which will help extend the
search region into the $\tanb \approx 10$ range as well as the $M_{H^\pm} >
M_{top}$ regime.

The Tevatron Higgs program promises to be an exciting field in the coming years 
with good discovery potential before the LHC turn-on. See Drew Baden's
talk\cite{baden} from this conference for a much more extensive overview of the
Tevatron's search potential in the next century.

\nonumsection{Acknowledgements}

\noindent I am indebted to the many people who provided me input for this talk. 
Thanks to Brendan Bevensee, John Hobbs, Bryan Lauer, Charles Loomis, Juan
Valls, Rocio Vilar and Weiming Yao for patiently explaining their analyses
and for their useful plots.

We thank the Fermilab staff and technical staffs of the participating
institutions for their vital contributions.

\nonumsection{References}

\end{document}